\documentclass[twocolumn,showpacs,prl,superscriptaddress,floatfix]{revtex4}

\RequirePackage{xspace}

\usepackage{dcolumn}
\usepackage{epsfig}

\input{pubboard/babarsym}

\def\Dmix    {\Dz-\Dzb \xspace}
\def\WSdec   {\ensuremath{\Dz\to\Kp\pim\piz} \xspace}
\def\xPrime  {\ensuremath{x^{\prime}_{K\pi\piz}}\xspace}
\def\xPrimeP {\ensuremath{x^{\prime +}_{K\pi\piz}}\xspace}
\def\xPrimeM {\ensuremath{x^{\prime -}_{K\pi\piz}}\xspace}
\def\yPrime  {\ensuremath{y^{\prime }_{K\pi\piz}}\xspace}
\def\yPrimeP {\ensuremath{y^{\prime +}_{K\pi\piz}}\xspace}
\def\yPrimeM {\ensuremath{y^{\prime -}_{K\pi\piz}}\xspace}
\def\dm      {\ensuremath{\Delta m}\xspace}
\def\mKpp    {\ensuremath{m_{K\pi\piz}}\xspace}
\def\tKpp    {\ensuremath{t_{K\pi\piz}}\xspace}

\newcommand{\BABARPubYear}    {08}
\newcommand{\BABARPubNumber}  {026}
\newcommand{\SLACPubNumber} {13320}

\begin{document}

\begin{flushleft}
\babar-PUB-\BABARPubYear/\BABARPubNumber\\
SLAC-PUB-\SLACPubNumber\\
\end{flushleft}

\title{\Large \bf \boldmath Measurement of \Dmix mixing from a time-dependent amplitude analysis of \WSdec\ decays}

%
\author{B.~Aubert}
\author{M.~Bona}
\author{Y.~Karyotakis}
\author{J.~P.~Lees}
\author{V.~Poireau}
\author{E.~Prencipe}
\author{X.~Prudent}
\author{V.~Tisserand}
\affiliation{Laboratoire de Physique des Particules, IN2P3/CNRS et Universit\'e de Savoie, F-74941 Annecy-Le-Vieux, France }
\author{J.~Garra~Tico}
\author{E.~Grauges}
\affiliation{Universitat de Barcelona, Facultat de Fisica, Departament ECM, E-08028 Barcelona, Spain }
% ...
\collaboration{The \babar\ Collaboration}
\noaffiliation

\date{\today}

\begin{abstract}
We present evidence of \Dmix mixing using a time-dependent amplitude analysis of the decay \WSdec in a data sample of 384\invfb\ collected with
the \babar\ detector at the \pep2 \epem collider at SLAC. Assuming \CP conservation, we measure the mixing parameters \xPrime $=$
[2.61 $\mbox{}^{\rm +0.57}_{\rm -0.68}$\,(stat.) $\pm$ 0.39\,(syst.)]\%, \yPrime $=$ [-0.06 $\mbox{}^{\rm +0.55}_{\rm -0.64}$\,(stat.) $\pm$ 0.34\,(syst.)]\%.
The confidence level for the data to be consistent with the no-mixing hypothesis is $0.1\%$, including systematic uncertainties. This result is inconsistent with
the no-mixing hypothesis with a significance of $3.2$ standard deviations. We find no evidence of \CP violation in mixing.
\end{abstract}


\maketitle
The mixing between the flavor eigenstates $|\Dz\rangle$ and $|\Dzb\rangle$ of the neutral $D$ meson depends upon the mass and width differences of the mass
eigenstates. In the standard model (SM), the \Dmix mixing contribution from short-distance effects is negligible~\cite{SMexp}.
This is due to Glashow-Iliopoulos-Maiani suppression of the first two quark generations and Cabibbo-Kobayashi-Maskawa suppression of the third.
Long-distance effects from intermediate states that couple to both \Dz and \Dzb dominate. Their contributions to the mixing parameters are difficult to predict,
but are estimated to be of the order $10^{-3}\textrm{--}10^{-2}$~\cite{SMexp}.
Several recent studies report evidence for mixing parameters at the $1\%$ level~\cite{expmixing}.
This is consistent with some SM expectations and provides constraints on
new physics models~\cite{NPexp}. If mixing occurs, the physical eigenstates $ |D_{1,2}\rangle = p|\Dz\rangle \pm q|\Dzb\rangle$ must have different masses
$M_{1,2}$ or widths $\Gamma_{1,2}$. Here $|p|^2 + |q|^2 = 1$. Mixing is parametrized using $x \equiv 2\mbox{ }(M_{1} - M_{2})/(\Gamma_{1} + \Gamma_{2})$ and
$y \equiv (\Gamma_{1} - \Gamma_{2})/(\Gamma_{1} + \Gamma_{2})$, where 1 (2) refers to the almost \CP-even (odd) eigenstate. If \CP is conserved, then $|p/q|=1$
and $\arg(q/p\cdot \bar{A_f}/A_f) = 0$. Here $A_f$ ($\bar{A_f}$) is the amplitude of the transition of the \Dz (\Dzb) to the final state $f$.

In this letter, we analyze events in which the flavor of the \Dz~\cite{note} is measured at production. We present the first time-dependent amplitude
analysis of the \WSdec Dalitz plot to extract the mixing parameters. Previously, we studied the time dependence of \WSdec decays integrated over large regions
of the Dalitz plot. We found no evidence for mixing~\cite{wilson}. However, certain regions of the phase space are more sensitive to mixing than others
(see below). Therefore the amplitude analysis is potentially more sensitive than our previous work. The wrong-sign (WS) decays \WSdec and the more copious
right-sign (RS) decays \Dz\to\Km\pip\piz are reconstructed. The RS decays proceed through a Cabibbo-favored (CF) amplitude. The WS decays proceed through a
coherent sum of a doubly Cabibbo-suppressed (DCS) amplitude and a CF amplitude produced by mixing. We identify RS and WS decays by reconstructing the
$\Dstarp \to \Dz\pi_s^{+}, \Dz \to K \pi \piz$  decay chain. The flavor of the \Dz candidate is determined from the charge of the low-momentum pion
($\pi_s^{+}$). The DCS and the CF amplitudes are described with isobar models~\cite{Muramatsu} as outlined below.

\begin{figure}[!ht]
 \includegraphics[width=0.48\linewidth]{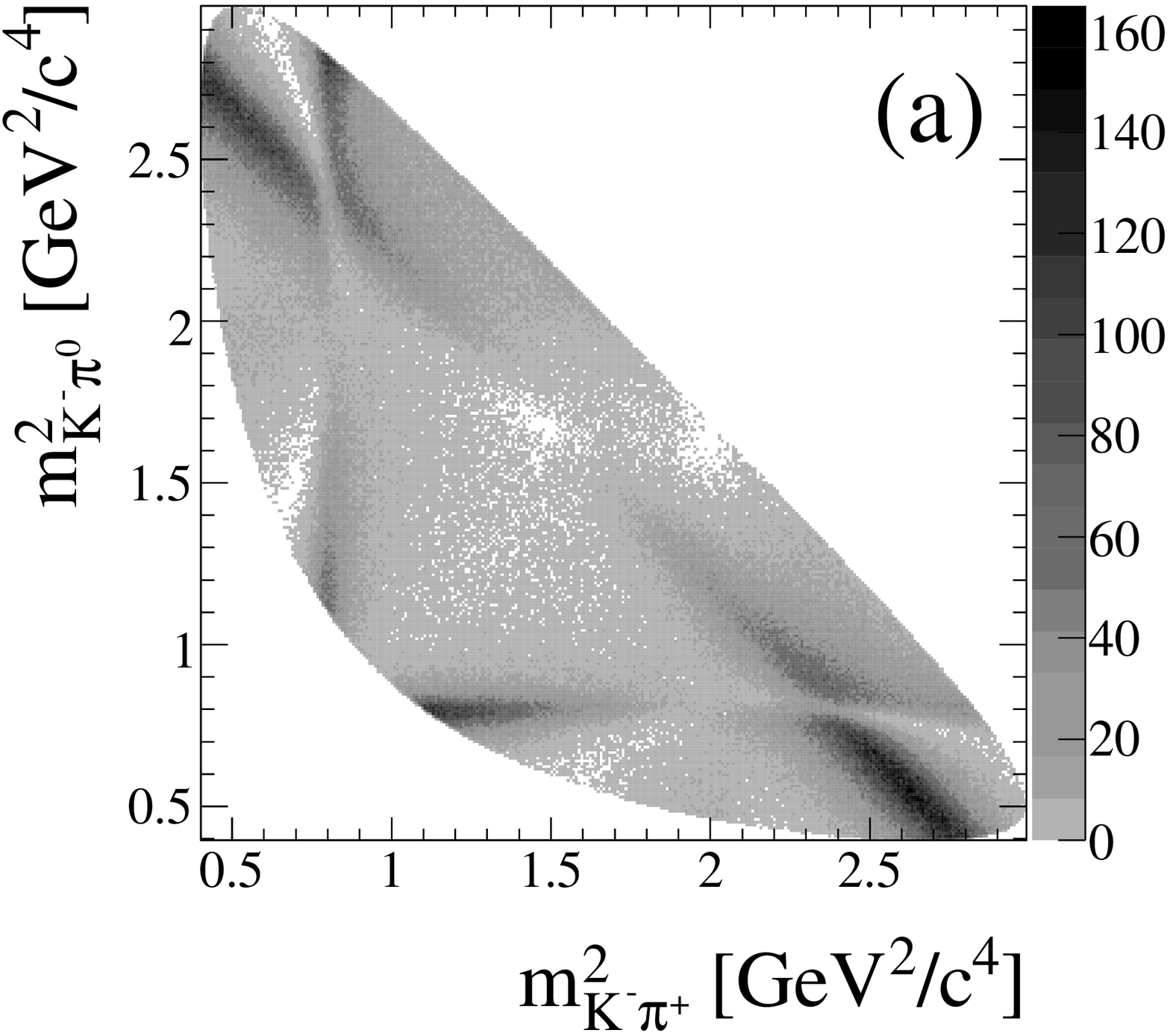}
 \includegraphics[width=0.48\linewidth]{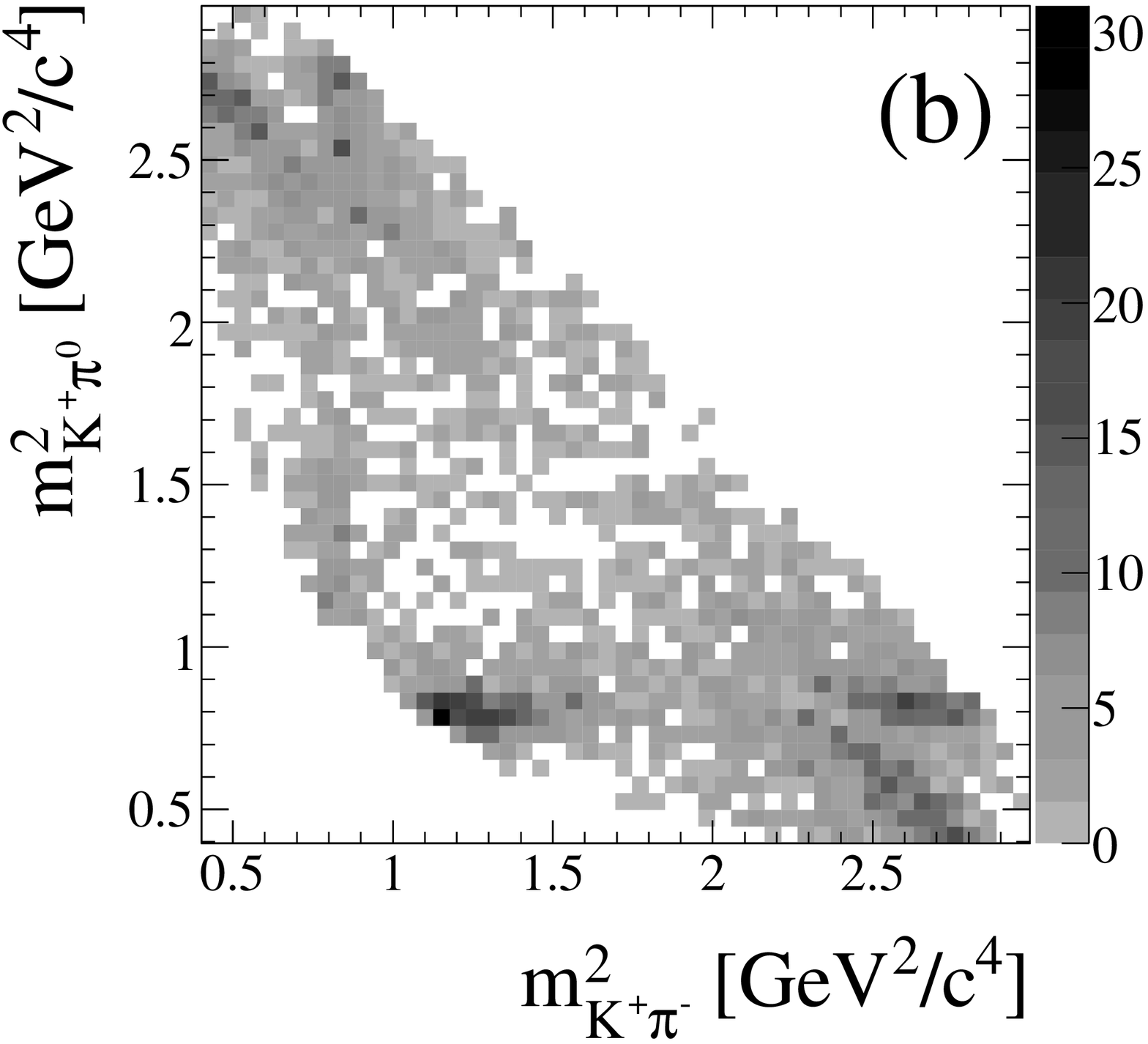}
 \includegraphics[width=0.48\linewidth]{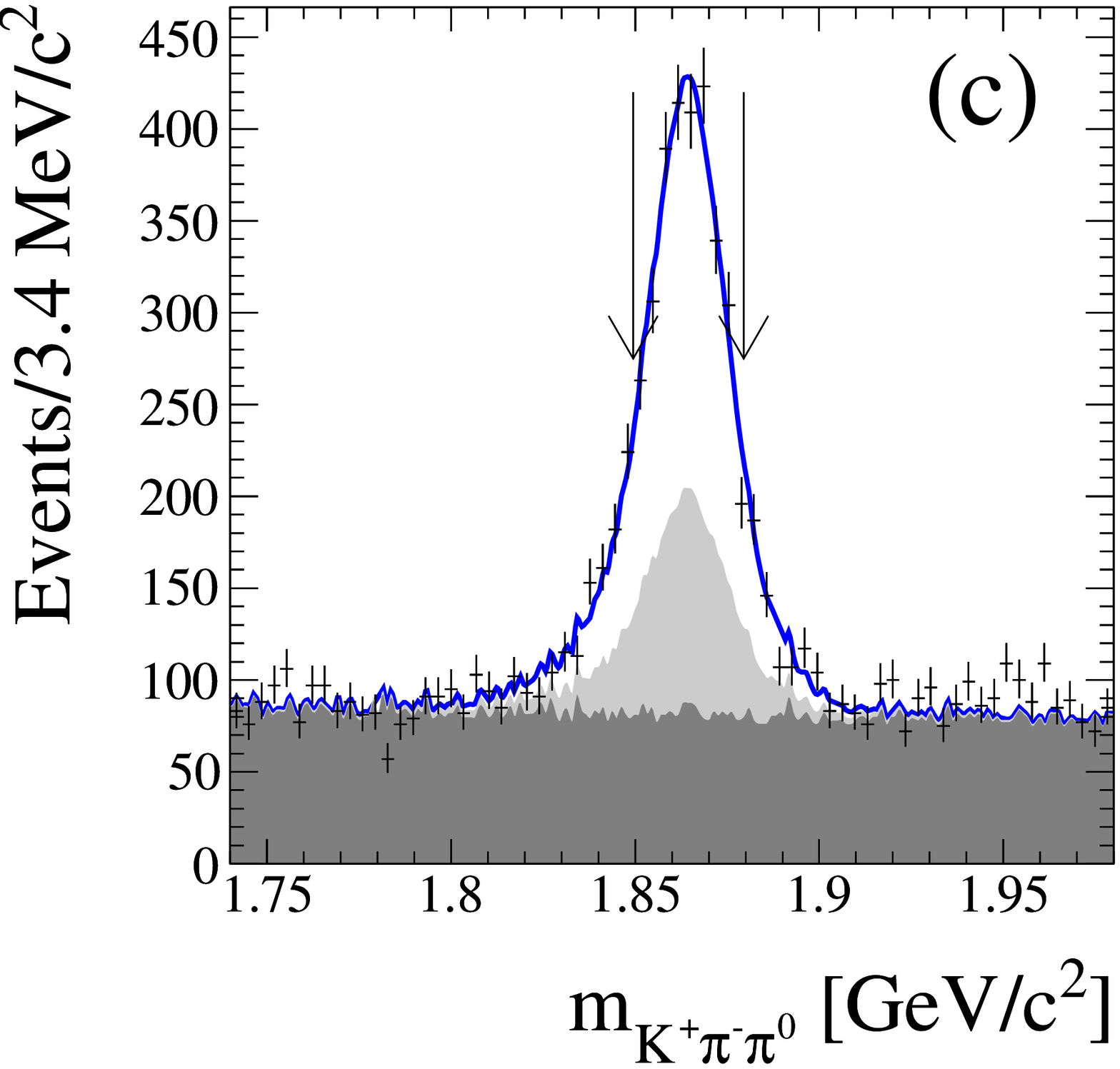}
 \includegraphics[width=0.48\linewidth]{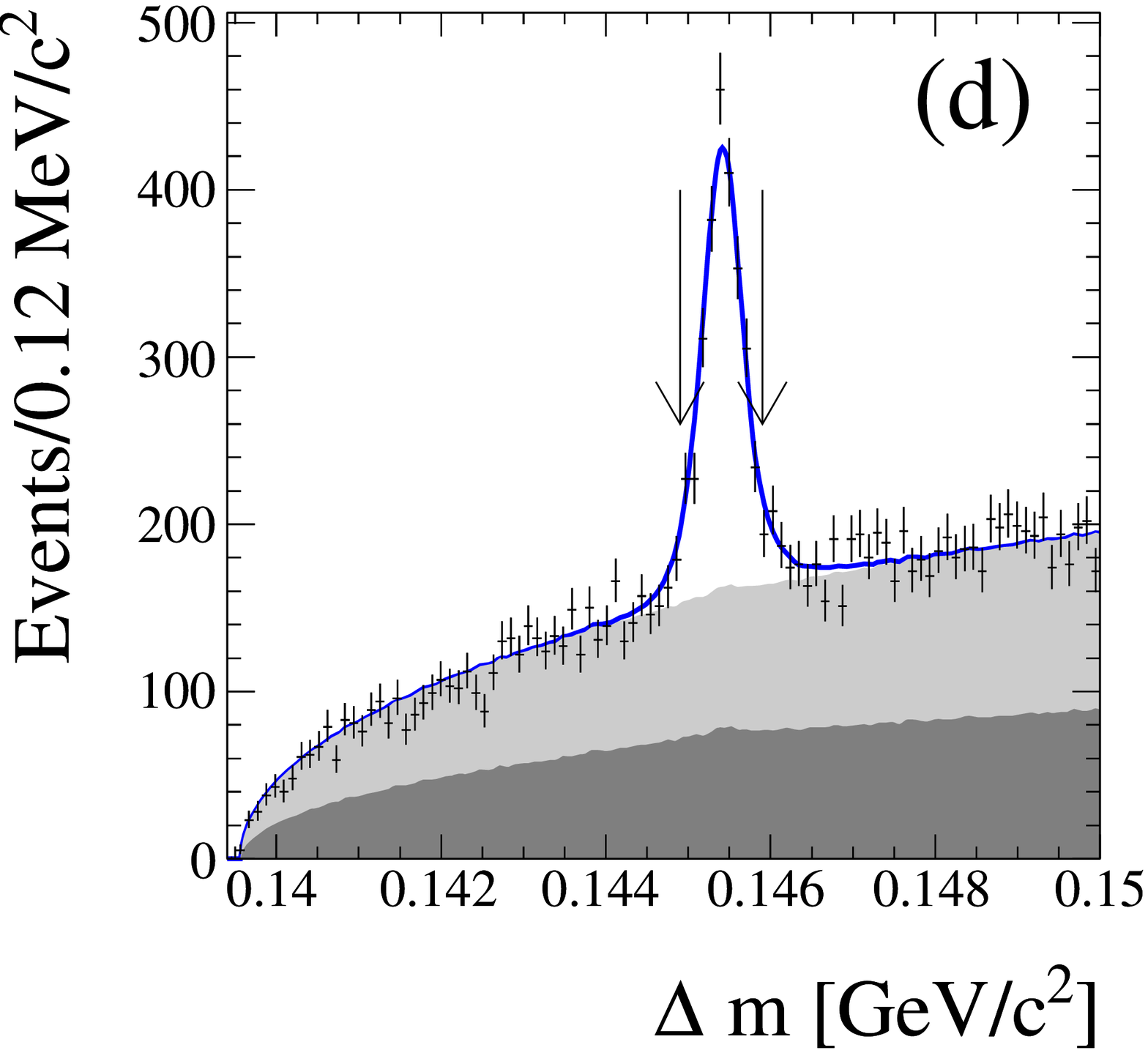}
\caption{\label{fig:plots1} Dalitz plots for the (a) RS and (b) WS \Dz samples. The reconstructed (c) \Dz mass and (d) \dm distributions for the WS sample
requiring respectively (c) $0.1449 < \dm\ < 0.1459$ \gevcc and (d) $1.8495 < \mKpp < 1.8795$ \gevcc. The fit results used to extract the yields are shown by the
superimposed curves. The light histogram represents the mistag background, while the dark histogram shows the combinatoric background.}
\end{figure}

The time-dependent decay rate is a function of the Dalitz variables $s_{12} = m^2_{\Kp\pim}$ and  $s_{13} = m^2_{\Kp\piz}$.
It depends on the DCS amplitude $A_{\bar{f}}(s_{12},s_{13}) = \langle\bar{f}|\mathcal{H}|\Dz\rangle $ and the CF amplitude
$\bar{A}_{\bar{f}}(s_{12},s_{13}) = \langle\bar{f}|\mathcal{H}|\Dzb\rangle$~\cite{SMexp}, where $\bar{f} = \Kp \pim \piz$. In the limit $|x|,|y| \ll 1$, it is
given by:
\begin{eqnarray}\label{eq:wsrate}
\frac{dN_{\bar{f}}( s_{12}, s_{13} ,t)}{ds_{12}ds_{13}dt} & = &  e^{-\Gamma t} \{  |A_{\bar{f}}|^2 + \nonumber \\
                  &   & |A_{\bar{f}}| |\bar{A}_{\bar{f}}| \left[ y\cos\delta_{\bar{f}} - x\sin\delta_{\bar{f}} \right] (\Gamma t) + \nonumber\\
                  &   &  \frac{x^{2}+y^{2}}{4} |\bar{A}_{\bar{f}}|^2 (\Gamma t)^2 \} 
\end{eqnarray}
where $\delta_{\bar{f}}( s_{12},s_{13}) = \arg[A_{\bar{f}}^{*}( s_{12},s_{13})\bar{A}_{\bar{f}}(s_{12},s_{13})]$. The first term in Eq.~\ref{eq:wsrate} is the DCS
contribution to the WS rate; the third term is a pure mixing contribution; the second term arises from the interference between DCS and mixing CF amplitudes.
We determine the CF amplitude $\bar{A}_{\bar{f}}$ in a time-independent Dalitz plot analysis of the RS decay sample, and use it in the analysis of the WS sample.
The DCS amplitude $A_{\bar{f}}$ is extracted along with the mixing parameters. In the isobar approach, $\bar{A}_{\bar{f}}$ and $A_{\bar{f}}$ are
described as a coherent sum of amplitudes, where each amplitude accounts for a resonance contribution. From inspection of the Dalitz plots
(Fig.~\ref{fig:plots1}a-b), WS decays proceed primarily through the resonance $\Dz\to \Kstarp\pim$, while RS  decays proceed primarily through
$\Dz\to \Km\rho^+$~\cite{Muramatsu}. For both $\bar{A}_{\bar{f}}$ and $A_{\bar{f}}$, one complex amplitude must be fixed arbitrarily.

The advantage of an amplitude analysis is that the interference term in Eq.~\ref{eq:wsrate} produces a variation in average decay time as a function of position
in the Dalitz plot that is sensitive to the complex amplitudes of the resonant isobars as well as the mixing parameters. In this study, the change in the
average decay time and the interference between the $\Dz\to \Kstarp\pim$ and $\Dz\to \rho^- \Kp$ amplitudes are the origin of our sensitivity to mixing. Our
analysis is sensitive to $x$ and $y$ in the form
$\yPrime \equiv y\cos{\delta_{K\pi\piz}} - x\sin{\delta_{K\pi\piz}}$ and $ \xPrime \equiv x\cos{\delta_{K\pi\piz}} + y\sin{\delta_{K\pi\piz}}$ where
$\delta_{K\pi\piz}$ is the strong interaction phase difference between the DCS $\Dz\rightarrow \rho^{-}\Kp$ and the CF $\Dzb\rightarrow \Kp\rho^{-}$
amplitudes, and cannot be determined in this analysis (note that $\delta_{K\pi\piz}$ is in general different from the analogous phase in other decays). 
Mixing is implied through a non-zero value of \xPrime or of \yPrime.
We define $A_{\bar{f}}( s_{12},s_{13})  =   r_{0} A_{\bar{f}}^{DCS}( s_{12},s_{13}) $ and
$\bar{A}_{\bar{f}}( s_{12},s_{13})  =  \bar{A}_{\bar{f}}^{CF}( s_{12},s_{13}) $. Here
\begin{eqnarray}\label{eq:r0def}
r_{0}^{2} = N_{WS}/\left(N_{RS}\cdot\int N_{\bar{f}}( s_{12}, s_{13} ,t)ds_{12}ds_{13}dt\right)
\end{eqnarray}
where $N_{WS}$ ($N_{RS}$) is the number of WS (RS) events in the sample. The variation of average decay time in the Dalitz plot due to mixing depends on the
ratios $\xPrime/r_0$ and $\yPrime/r_0$; these are the parameters that we extract directly from the data and use to determine the significance of the mixing
result.

The amplitudes entering the WS analysis are described as a sum of isobar components $A_j$ that are parametrized with Breit-Wigner functions,
$ A_{\bar{f}}^{CF/DCS} = \sum_{j=1}^{n_{CF/DCS}} a_j e^{i\delta_j} A_j( m^2_{\Kp\pim}, m^2_{\Kp\piz})$, where $a_{j}$ and $\delta_{j}$ are the strong interaction
amplitudes and phases of the $j^{th}$ resonant amplitude~\cite{Muramatsu}. For the $K$-$\pi$ S-wave component we use a parametrization derived from $K$-$\pi$
scattering data~\cite{lass}, which consists of a $\Kstar_{0}(1430)$ resonance together with an effective non-resonant component.

We analyze a data sample of $384$ \invfb collected with the \babar\ detector~\cite{Aubert:2001tu} at the \pep2 \epem collider at SLAC near a center-of-mass
energy of $10.58$ \gev. Charged tracks are reconstructed with a silicon-strip detector (SVT) and a drift chamber (DCH), both in a 1.5\,T magnetic field. Particle
identification is based on measurements of ionization energy loss (\dedx) in the SVT and DCH together with measurements from a Cherenkov ring-imaging device.
Photon energies are measured with a CsI(Tl) calorimeter. All selection criteria, the fit procedure and the systematic error analysis are finalized before we
search for evidence of mixing in the data.

Selection criteria are based partly on those of Ref.\cite{wilson} and are identical for the RS and WS samples. We require the $\pi_s^{+}$ candidates to have a
transverse momentum $p_t^{LAB}> 0.12$\gevc, where $LAB$ indicates the laboratory frame. We reject electrons that mimic $\pi_s^{+}$ using \dedx measurements. We
use kinematic selection criteria to eliminate electrons from pair conversions. The energies of photon candidates used to form \piz candidates are required to be
greater than $0.1$\gev; the invariant mass of photon pairs forming a \piz must be in the range $0.09  < m_{\piz}  <  0.16$ \gevcc. We require the \piz momentum
$p_{\piz}^{LAB}$ to
be greater than $0.35$\gevc . The reconstructed invariant mass for the \Dz candidates must have  $1.74 < \mKpp < 1.98\gevcc$. The \piz and \Dz masses are
then set equal to their nominal values \cite{pdg2006} and the \Dstar is refitted \cite{Hulsbergen:2005pu} with the constraint that its production point lies
within the beam spot region. The \Dstarp invariant  mass and \Dz measured decay time \tKpp are derived from this fit. We require $0.139 < \dm < 0.155\gevcc$ where
$\dm \equiv m_{K\pi\piz\pi_{s}} - \mKpp$. To reject \Dstar candidates from $B$ decays, we require the \Dz center-of-mass momentum to be greater than $2.4$\gevc.
For events that contain multiple \Dstar candidates with shared tracks, the candidate that yields the largest fit probability for the decay chain is retained.
The three-dimensional flight path is used to determine \tKpp and its uncertainty $\sigma_t$. For signal events, the typical value of $\sigma_t$ is 0.23\ps;
we accept \Dstar candidates with $\sigma_t < 0.50$ ps.

We extract the signal and background yields from a binned extended maximum likelihood fit to the \mKpp and \dm distributions (Fig.~\ref{fig:plots1}c-d).
For subsequent analysis, we retain \Dstar candidates in the signal region defined as $0.1449 < \dm\ < 0.1459$ \gevcc and $1.8495 < \mKpp\ < 1.8795$ \gevcc .
Our final RS (WS) sample is composed of $658,986$ ($3009$) events with a purity of $99\%$ ($50\%$). The efficiency of the signal region selection is $54.6\%$.

\begin{figure}[!ht]
\includegraphics[width=0.48\linewidth]{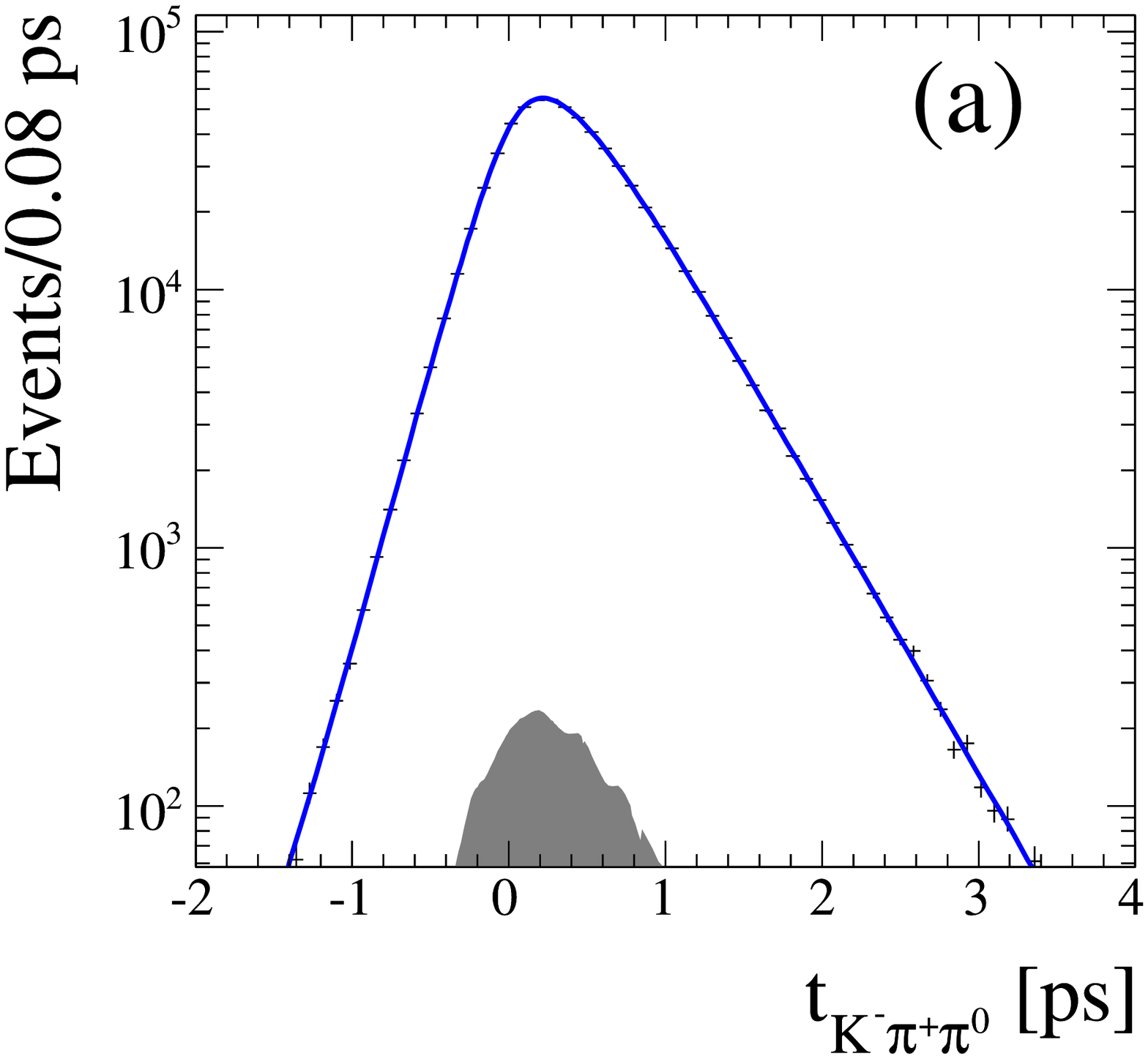}
\includegraphics[width=0.48\linewidth]{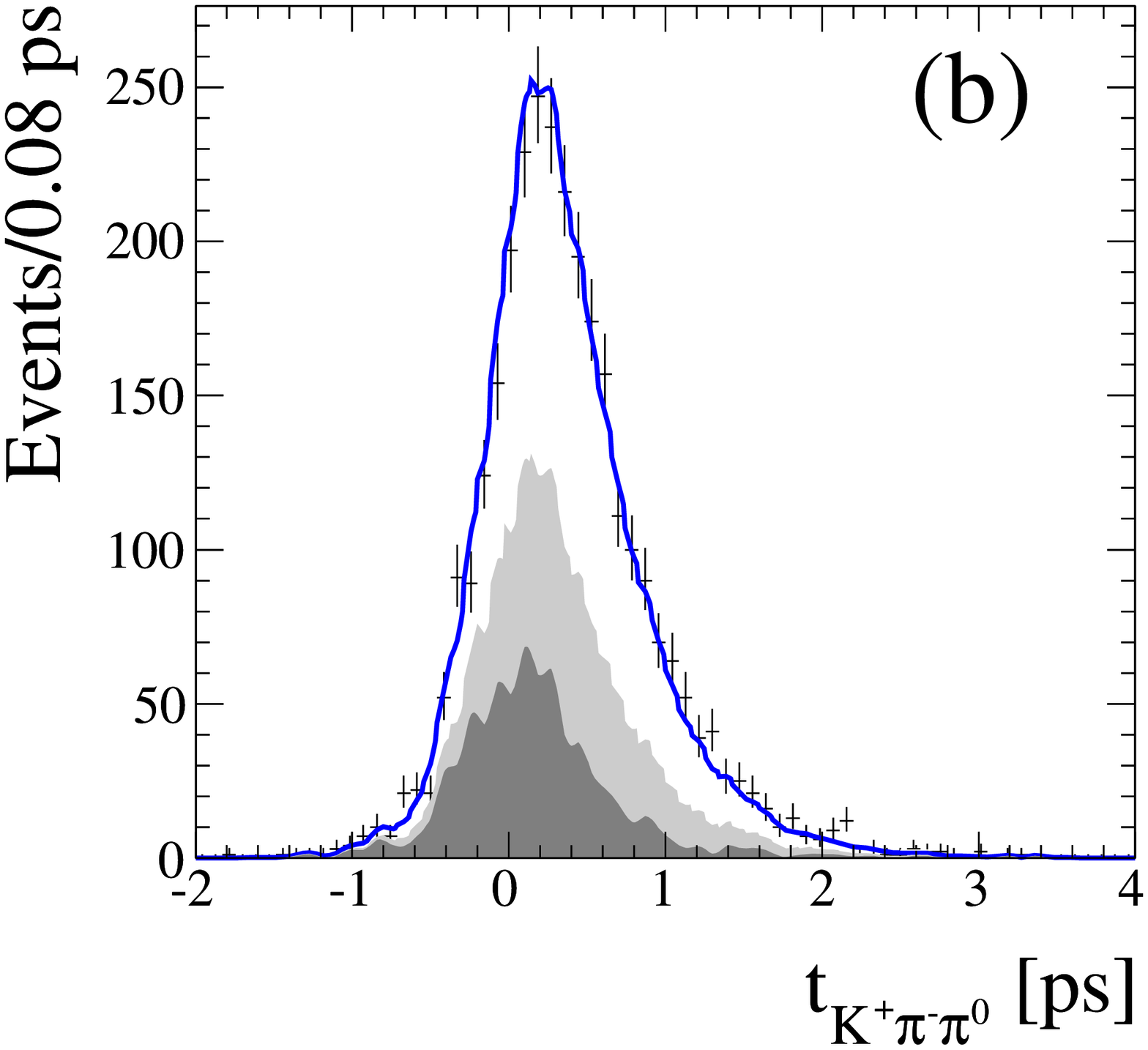}
\includegraphics[width=0.48\linewidth]{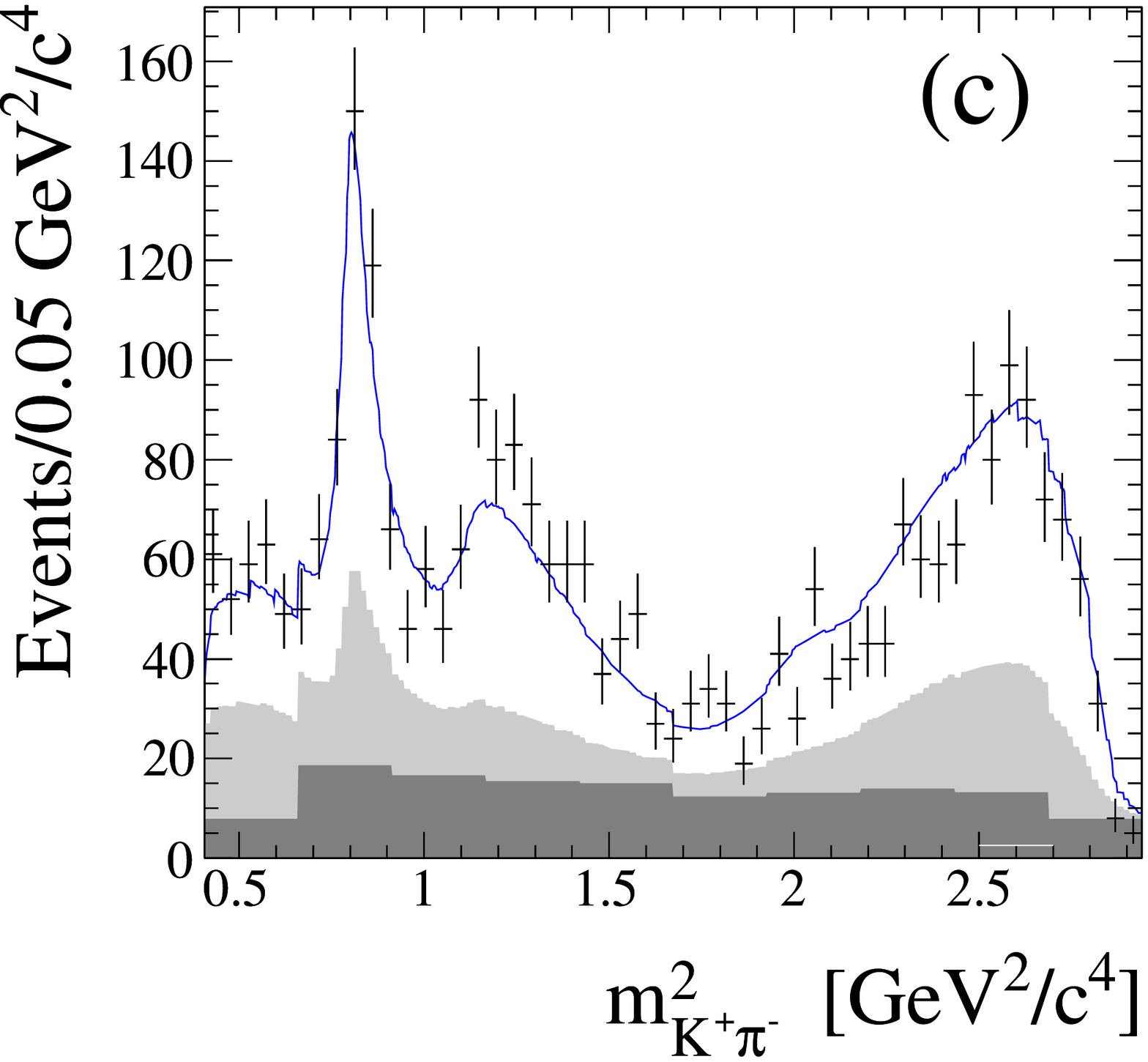}
\includegraphics[width=0.48\linewidth]{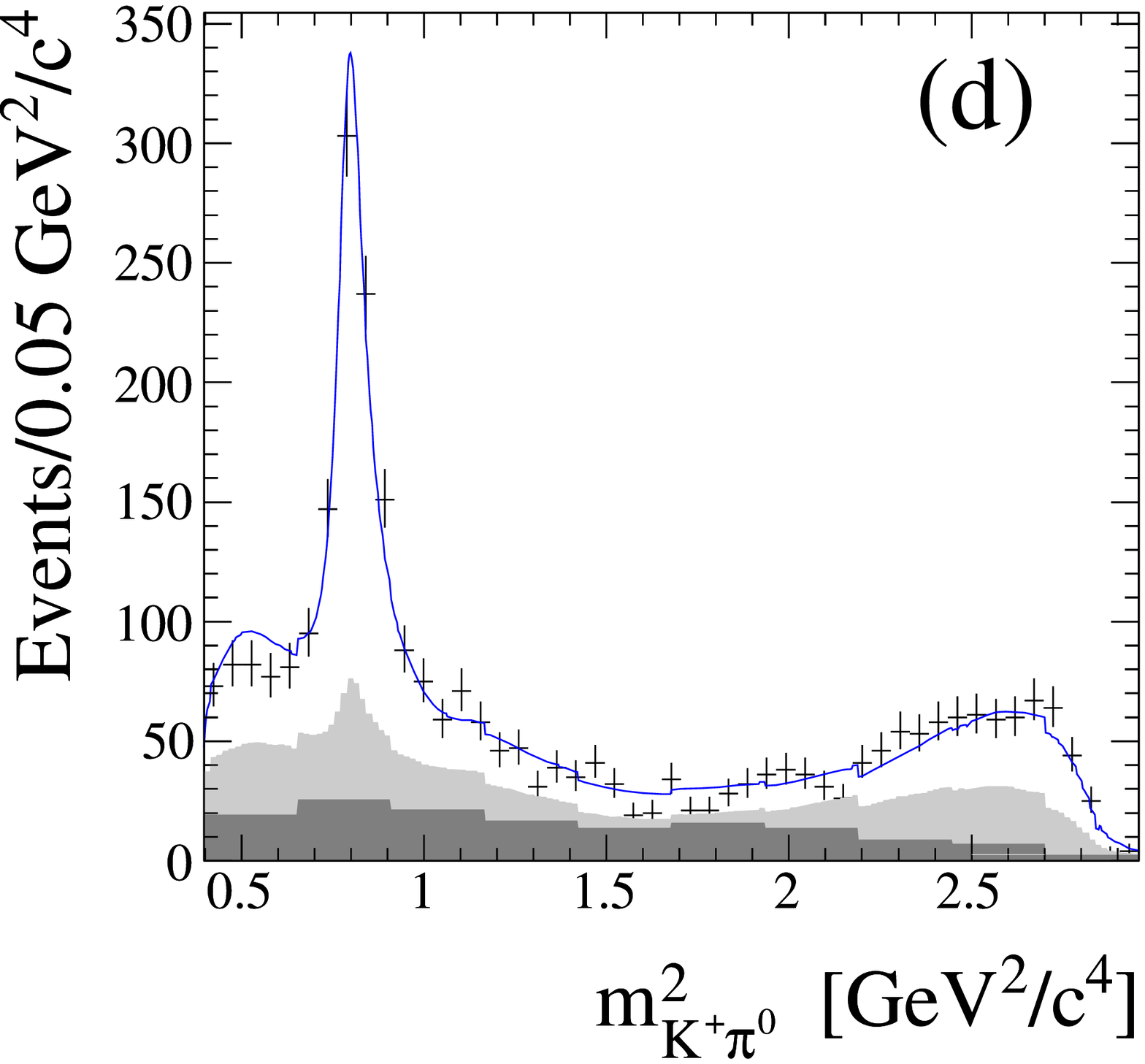}
\caption{\label{fig:plots2} (a) Proper time distribution for RS events with the fit result superimposed. The distribution of background events is shown by the
shaded histogram. (b) Proper time distribution for WS events. (c, d) $m^2_{\Kp\pim}$ and $m^2_{\Kp\piz}$ projections with superimposed fit results (line). The
light histogram represents the mistag background, while the dark histogram shows the combinatoric background;}
\end{figure}

The RS sample is used to determine the CF isobar model parameters $a^{CF}_j$ and $\delta^{CF}_j$, as well as the decay time resolution function, which is
parametrized as a sum of three Gaussian functions with a common mean, with widths given by the per-event $\sigma_t$ times a different scale factor for each
Gaussian. The reconstructed RS \Dz signal decay time distribution (Fig.~\ref{fig:plots2}a) is described by a probability density function (PDF) consisting of an
exponential function convolved with the resolution function. The resolution function parameters and \Dz lifetime are determined in an unbinned maximum likelihood
fit. The mean value of the resolution function is found to be $4.2 \pm 0.7$~fs. This value is consistent with the magnitude expected from instrumental effects,
and the associated systematic uncertainty is determined by setting the value to zero.
As a cross-check we determine the \Dz mean lifetime to be [$409.9 \pm 0.8$ (stat.\ only)]\,fs, in agreement with the world average
[$410.1 \pm 1.5$ (stat.\ +\ syst.)]\,fs \cite{pdg2006}.

The \Dz candidates in the WS signal region can be divided into three categories: signal events, combinatorial background, and incorrectly tagged RS events
(mistag), each one described by its own PDF, whose parameters are determined in an unbinned maximum likelihood fit. During the fit procedure, the number
of events in each category is fixed to the value obtained from the fit to the \mKpp and \dm distributions.

The PDF describing the WS time-dependent Dalitz plot is given by Eq.~\ref{eq:wsrate} convolved with the \tKpp resolution function. The $\sigma_t$ PDFs for
signal and background are taken from the RS data. The DCS amplitudes and phases
for each resonance and the mixing parameters are determined in the fit. The CF Dalitz plot amplitudes arising from mixing are taken from the fit to the RS sample
described in the previous paragraph. The mistag events are parametrized using an empirical PDF obtained from the RS data, since mistag events contain correctly
reconstructed RS \Dz decays. The PDF describing the combinatorial background is constructed by averaging the ($s_{12},s_{13},\tKpp$) distributions obtained from
the WS \mKpp sidebands: this accounts for correlations between those three variables that might be present in the data.

The results of the time-dependent fit of the WS data, the $a^{DCS}_j$, $\delta^{DCS}_j$ and fit fractions $f_j$ \cite{Muramatsu}, are given in
Table~\ref{tbl:results}.
The fit fraction of the non-resonant contribution to the $K$-$\pi$ S-wave is absorbed into the $\Kstarp_{0} (1430)$ and $\Kstarz_{0} (1430)$ fit fractions.
Projections of the fit results are shown in Fig.~\ref{fig:plots2}b-d.
The change in log likelihood ($-2\Delta\ln{\cal L}$) between the fit with mixing and with no mixing ($\xPrime/r_0  = \yPrime/r_0  = 0$) is $13.5$ units, including
systematic uncertainties. For two degrees of freedom, the confidence level that the result is due to no-mixing is $0.1\%$. The significance of the mixing result
is equivalent to $3.2$ standard deviations, and thus constitutes evidence for \Dmix mixing.
\begin{table}[htbp]
\caption{Fit results for the WS \Dz data sample. The total fit fraction is $102\%$ and the $\chi^{2}/ndof$ is $188/215$. The results for $\xPrime/r_0$ and
$\yPrime/r_0$ include statistical and systematic errors; their total linear correlation is $-0.34$.}
\begin{center}
\begin{tabular}{|l|c|c|c|}
\hline
Resonance & $a^{DCS}_j$ & $\delta^{DCS}_j$ (degrees) & $f_j$ (\%)  \\
\hline
$\rho (770)$         &  1 (fixed)           &  0 (fixed)         & $39.8\pm 6.5$ \\
$\Kstarz_{2}(1430)$  & $  0.088\pm 0.017$   &  $ -17.2\pm 12.9$  & $2.0\pm 0.7$  \\
$\Kstarp_{0} (1430)$ & $  6.78\pm 1.00$     &  $ 69.1\pm 10.9$   & $13.1\pm 3.3$ \\
$\Kstarp (892)$      & $  0.899\pm 0.005$   &  $ -171.0\pm 5.9$  & $35.6\pm 5.5$ \\
$\Kstarz_{0} (1430)$ & $  1.65\pm 0.59$     &  $ -44.4\pm 18.5$  & $2.8\pm 1.5$  \\
$\Kstarz (892)$      & $  0.398\pm 0.038$   &  $ 24.1\pm 9.8$    & $6.5\pm 1.4$  \\
$\rho (1700) $       & $  5.4\pm 1.6$       &  $ 157.4\pm 20.3$  & $2.0\pm 1.1$  \\
\hline\hline
\multicolumn{4}{|c|}{$\xPrime/r_0 = 0.353\pm 0.091 \pm 0.052$ } \\
\multicolumn{4}{|c|}{$\yPrime/r_0 = -0.002\pm 0.090 \pm 0.059$ }\\
\hline
\end{tabular}
\end{center}
\label{tbl:results}
\end{table}

To derive the values of $\xPrime$ and $\yPrime$ we first determine $r^2_0=$ [$5.25^{+0.25}_{-0.31}$ (stat.) $\pm 0.12$ (syst.)] $\times 10^{-3}$ using
Eq.~\ref{eq:r0def}. We then generate $10^6$ ($\xPrime/r_0$,$\yPrime/r_0$) points in accordance with the fit covariance matrix, assuming Gaussian errors
(width given by the total uncertainty including systematics). For each point, we compute $r_0$ using Eq.~\ref{eq:r0def} and determine values for \xPrime and
\yPrime. Using a Bayesian approach, by integrating the likelihood function with respect to \xPrime and \yPrime, assuming a flat prior distribution, we obtain
\xPrime=[$2.61^{+0.57}_{-0.68}$ (stat.) $\pm 0.39$ (syst.)]\% and \yPrime = [$-0.06 ^{+0.55}_{-0.64}$ (stat.) $\pm 0.34$ (syst.)]\% with a correlation of $-0.75$.

Extensive validation of this fitting procedure is performed using Monte Carlo (MC) experiments based on the PDF shapes and DCS amplitudes extracted from data.
The validation studies are performed over a wide range of mixing parameters. These studies demonstrate that the fit correctly determines the mixing parameters to
within a small offset of $0.2$-$0.3\sigma$, where $\sigma$ is the statistical uncertainty. These small biases are a consequence of the relatively small size of
our data sample and become negligible if MC samples with higher statistics are used. We correct the final result for this offset.

Sources of systematic uncertainty for $\xPrime/r_0$($\yPrime/r_0$), related to the choice of the isobar model and the experimental assumptions, are considered.
For each effect we refit the data with an alternative assumption and extract the overall  correlated uncertainty for the fitted parameters.
We estimate the Dalitz model uncertainties [$0.38\sigma$ ($0.35\sigma$)], where $\sigma$ is the statistical uncertainty, by varying the mass and the width of each
resonance within their error and by using alternative parametrizations for the isobar components $A_j$ in the fit: the largest error arises from uncertainties in
the $\Kstar$ and $\rho$ parameters and from uncertainties in the parametrization of the $K$-$\pi$ S-wave. Systematic uncertainties related to the number
of signal and background events [$0.15\sigma$ ($0.22\sigma$)] are evaluated by varying them according to their statistical uncertainties. Similarly, the
definition of the signal region, the $\sigma_t$ requirement, and the selection of the best \Dstar candidate are varied. The effect on the mixing parameters is
$0.50\sigma$ ($0.37\sigma$). Variations in efficiency across the Dalitz plot contribute systematic uncertainties of $0.09\sigma$ ($0.10\sigma$). The \tKpp
resolution function parameters are varied within their errors. The offset is also set to zero. The systematic effect is $0.11\sigma$ ($0.09\sigma$).
The total systematic error on $\xPrime/r_0$ ($\yPrime/r_0$) is $0.57\sigma$ ($0.66\sigma$).

The same procedure is applied separately to the WS \Dz-tagged ($+$) and \Dzb-tagged ($-$) events to search for \CP violation in  mixing. We find
$\xPrimeP = (2.53 ^{+0.54}_{-0.63}\pm 0.39)\%$, $\yPrimeP = (-0.05 ^{+0.63}_{-0.67}\pm 0.50)\%$, $\xPrimeM = (3.55 ^{+0.73}_{-0.83}\pm 0.65)\% $ and
$\yPrimeM  = (-0.54 ^{+0.40}_{-1.16}\pm 0.41)\%$, respectively, and thus observe no evidence for \CP violation. The correlation between \xPrimeP (\xPrimeM) and
\yPrimeP (\yPrimeM) is $-0.69$ ($-0.66$).  

In conclusion, our data are inconsistent with the no-mixing hypothesis with a significance of $3.2$ standard deviations.
Our results thus constitute evidence for mixing. For the mixing parameters we find
$\xPrime = (2.61^{+0.57}_{-0.68} \pm 0.39)\%$ and $\yPrime =(-0.06 ^{+0.55}_{-0.64}\pm 0.34)\%$ with a correlation of $-0.75$. These values are consistent with
our previous result~\cite{wilson} and with some SM estimates for mixing. No evidence for \CP violation is found. 

We are grateful for the excellent luminosity and machine conditions
provided by our \pep2\ colleagues, 
and for the substantial dedicated effort from
the computing organizations that support \babar.
The collaborating institutions wish to thank 
SLAC for its support and kind hospitality. 
This work is supported by
DOE
and NSF (USA),
NSERC (Canada),
CEA and
CNRS-IN2P3
(France),
BMBF and DFG
(Germany),
INFN (Italy),
FOM (The Netherlands),
NFR (Norway),
MES (Russia),
MEC (Spain), and
STFC (United Kingdom). 
Individuals have received support from the
Marie Curie EIF (European Union) and
the A.~P.~Sloan Foundation.


\begin{thebibliography}{99}
\bibitem{note}
{The use of charge-conjugate modes is implied unless otherwise noted.}
\end{thebibliography}
\end{document}